\documentclass[prb, twocolumn,showpacs, superscriptaddress]{revtex4}
\usepackage{graphicx}
\usepackage{dcolumn}

\begin{document}
\title{Scattering of carriers by charged dislocations in semiconductors}
\author{Bhavtosh Bansal}\email{bhavtosh@iiserkol.ac.in}
\affiliation{Indian Institute of Science Education \& Research, Kolkata, Mohanpur Campus, Nadia 741252, West Bengal, India}
\author{Rituparna Ghosh}
\affiliation{Indian Institute of Science Education \& Research, Kolkata, Mohanpur Campus, Nadia 741252, West Bengal, India}
\author{V. Venkataraman}
\affiliation{Department of Physics, Indian Institute of Science, Bangalore 560 012, India}
\date{\today}
\begin{abstract}
The scattering of carriers by charged dislocations in semiconductors is studied within the framework of the linearized Boltzmann transport theory with an emphasis on examining consequences of the extreme anisotropy of the scattering potential. A new closed-form approximate expression for the carrier mobility valid for all temperatures is proposed. The ratios of quantum and transport scattering times are evaluated after averaging over the anisotropy in the relaxation time. The value of the Hall scattering factor computed for charged dislocation scattering indicates that there may be a factor of two error in the experimental mobility estimates using the Hall data. An expression for the resistivity tensor when the dislocations are tilted with respect to the plane of transport is derived. Finally an expression for the isotropic relaxation time is derived when the dislocations are located within the sample with a uniform angular distribution.
\end{abstract}
\pacs{72.10.Fk, 72.20.Dp}
\maketitle
\section{Introduction}
Epitaxial growth of thin semiconductor films on substrates which have a large lattice constant mismatch results in the films being strained. Depending on the growth conditions and the films' thickness, this strain can either partially or fully relax through a formation of various possible kinds of lattice defects. Among these defects edge dislocations are prominent and have a pronounced effect on the mobility of carriers.\cite{SolidStatePhysics}
While the theory for charged dislocation scattering was first formulated to explain the low temperature mobility of plastically deformed semiconductors\cite{podor}, interest in dislocation scattering has revived in the last 15 years in context of  GaN \cite{{SolidStatePhysics}, look,mse-review, GaN1,farvacque-prb} and InN\cite{InN1} which typically do not have lattice-matched substrates. Indeed it is important in all epitaxially grown materials \cite{3-5_on_Si, LaAlO3} on mismatched substrates, as well as bulk crystals whose growth techniques have not yet been mastered.\cite{bulk-apl}

An edge dislocation is a row of dangling bonds formed by an abruptly terminated plane somewhere inside the crystal.\cite{mse-review} This local departure from tetragonal coordination produces acceptor states in the energy gap, forming one dimensional lines of charge. The effective screened electrostatic potential energy, $U(x_\perp)$ is thus cylindrically symmetric if the extent of the edge dislocation is taken to be infinite.\cite{SolidStatePhysics, seeger,handbook}
\begin{equation}\label{real space potential}
U(x_{_\perp})={Qe\over 2\pi\epsilon}K_0(x_{_\perp}/\lambda)
\end{equation}
where $Q$ is the charge per unit length, $K_0$ is the modified zeroth order Bessel function of the second kind, $\epsilon=\epsilon_0 \epsilon_r$ is the dielectric constant, $\lambda$ is the screening length and $x_{_\perp}$ is the distance from the dislocation line in a perpendicular plane, ${\bf r}=f(x_{_\perp}, \theta,z)$. These one dimensional lines of charge have detrimental effects on the transport properties of charge carriers.

\section{Issues Addressed}
\noindent
(i)  The scattering potential $[$Eq. (1)$]$ is highly anisotropic due to its cylindrical symmetry. It is known that the relaxation time approximation for the solution of the linearized Boltzmann
equation is in general not valid for anisotropic potentials. In context of the charged dislocation scattering also, the extension of the relaxation time approach has been questioned.\cite{{mse-review}, farvacque-prb, farvacque_sst} We will, first of all, rigorously establish the existence of a relaxation time for this problem.\\
(ii) We will next show that P\"{o}d\"{o}r's expression for the relaxation time\cite{podor}
\begin{equation}\label{podor's expression}
\tau={8\epsilon^2\, m^{\ast 2}\over N_d e^2 Q^2 \lambda}({\hbar^2\over 4m^{\ast %%@
2}\lambda^2}+v_{\perp}^2)^{3/2}
\end{equation}
is indeed correct, despite an apparent inconsistency. In Eq. (\ref{podor's expression}), $v_{\perp}$ is the
component of electron velocity perpendicular to the dislocation
axis and $N_d$ is the number of dislocations per unit area, all
assumed to be parallel and independent. Only the
perpendicular component of the impinging electron's velocity contributes to scattering and the component parallel to the dislocation is unaffected. Eq. (\ref{podor's expression}) is finite when $v_{\perp}\rightarrow 0$, whereas in this limit, $\tau$ should diverge. This point gets clarified once one goes through a consistent derivation of the relation time in the next section where we break up the relaxation time into two components $\tau_\perp$ and $\tau_z$. $\tau_\perp$, the component of the relaxation time perpendicular to dislocations does indeed correspond to P\"{o}d\"{o}r's expression whereas $\tau_z$, the component parallel to the dislocation axes  is ill-defined.\\
(iii) The method of energy averaging employed by P\"od\"or has been questioned.\cite{look} Due to this ambiguity, the tensor nature of resistivity is not evident in the final expression. In particular if the dislocations are tilted at an angle with respect to the direction perpendicular to the plane of transport, it is difficult to give anything better than a rough estimate in the present theory.\cite{Eastman} The effect of dislocation orientation is usually disregarded and $\mu_\perp$ is replaced by a scalar number.\cite{SolidStatePhysics} Nevertheless dislocation related anisotropy is sometimes seen in the transport properties.\cite{anisotropy}\\
(iv) Quantum and classical scattering times were calculated without averaging out the anisotropy in the problem.\cite{jena}\\
(v) There are corrections to the measured Hall mobility due to the Hall scattering factor. This Hall factor is shown to be very significant, even larger than 2 for a non-degenerate electron gas.\\
(vi) In general, dislocations may not be all parallel. A naive angular averaging over the resistivity tensor is equivalent to the use of Matthiessen's rule. We will derive a new expression for angular-averaged relation time $\tau_{iso}(k)$, which has a different $k$ dependence as compared to the anisotropic relaxation time. Thus angular averaging has the important experimental consequence of changing the temperature dependence of mobility.
\section{Theoretical Formulation}
Let us start from the Boltzmann equation within the linear response regime.\cite{ziman}
Then up to the first order in electric field, the perturbed distribution function may symbolically be written as $f_{\bf k}=f_{0{k}}+\phi_{\bf k}$, where $\phi_{\bf k}$ is deviation from an equilibrium distribution in presence of a perturbing external electric field ${\bf F}$.
In absence of a thermal gradient and a magnetic field, the linearized Boltzmann equation for carriers described by spherical parabolic band reduces to
\begin{equation} \label{linear-boltzmann}
{e\hbar\over m^{\ast}}{\bf F\cdot k}{\partial f_{0k}\over \partial E}=\sum_{\bf %%@
k^\prime}W_{\bf k^\prime,k }[\phi_{{\bf k}^\prime}-\phi_{{\bf k}}]
\end{equation}
$W_{\bf k, k^{\prime}}$ is the transition rate between initial and final plane %%@
wave states,  ${\bf k }$ and ${\bf k^{\prime}}$, in presence of the scattering %%@
potential given by Eq. (\ref{real space potential}). For scattering from %%@
charged dislocations, the scattering rate is given by $W_{\bf k, k^{\prime}} = %%@
\delta(k_z-k^\prime_z)\delta(k-k^\prime)g(|{\bf k}^\prime_\perp-{\bf k}_\perp|)$. %%@
$g(|{\bf k}^\prime_\perp-{\bf k}_\perp|)$ is the part depending on only a function %%@
of in-plane momenta (shown below). Thus (a) collisions are elastic, (b) the %%@
components of the incident electron's momenta which are parallel and perpendicular %%@
to the dislocation line are separately conserved, (c) no electric field develops %%@
along the dislocation axis, i.e. ${\bf F\cdot k}={\bf F}_\perp\cdot{\bf k}_\perp$. %%@
This immediately implies that no relaxation time can be defined along the %%@
direction parallel to the dislocations' axis. In other words, for time independent %%@
electric field, there is no steady state solution to the Boltzmann equation if the %%@
collision term is zero. Nevertheless, one may physically argue that %%@
$1/\tau_{z}=0$. The argument is clear within the variational formalism %%@
where one defines the sample resistivity in terms of the Joule-heat dissipated due %%@
to a finite current (see appendix).\cite{ziman} With constraints (a)-(c) in mind, we shall choose a %%@
$\phi_{\bf k}$ which solves the linearized Boltzmann's equation {\em exactly}. Ansatz:
\begin{equation}\label{ansatz}
\phi_{\bf k}= -{e\hbar\over m^{\ast}}\tau_\perp(k_\perp){\bf F_\perp\cdot %%@
k_\perp}{\partial f_{0k} \over \partial E}
\end{equation}
Substituting $\phi_{\bf k}$ in Eq. (\ref{linear-boltzmann}) yields
\begin{eqnarray}
{\partial f_{0k}\over \partial E} {\bf F}_\perp\cdot{\bf k}_\perp =
{\bf F}_\perp\cdot \sum_{{\bf k}^\prime} {W_{\bf k, %%@
k^\prime}}\hspace{2.5cm}\nonumber\\ \times\left[
{\partial f_{0k}\over \partial E}\tau_\perp(k_\perp){\bf %%@
k}_\perp-{\partial f_{0k^\prime}\over \partial
E}\tau_\perp(k^\prime_\perp){{\bf k}^\prime}_\perp\right]
\end{eqnarray}
From energy and perpendicular momentum conservation,
\begin{equation}
{\partial f_{0k^\prime} \over \partial E}\tau_\perp(k^\prime_\perp)={\partial %%@
f_{0k} \over \partial E}\tau_\perp(k_\perp).\end{equation} Thus the linearized %%@
Boltzmann equation is exactly solved if
\begin{equation}
{1\over \tau_\perp(k_\perp)}=\sum_{{\bf k}^\prime} W_{\bf k, k^\prime}(1-cos\,\theta) %%@
\label{relaxdef}
\end{equation}
Here $\theta$ is the angle between ${\bf k}_\perp, {\bf k^\prime}_\perp$ which lie %%@
on a circle parallel to the xy-plane since $k_z$ is independently
conserved. The wave vectors in the summation in Eq. (\ref{relaxdef}) are %%@
three dimensional.

Within the Born approximation
\begin{eqnarray} \label{w- k kprime}
W_{{\bf k},{\bf k}^{\prime}}={2\pi\over \hbar} \delta(E_k-E_{k^\prime})\left[{1\over L_xL_yL_z}\int d{\bf x}\,U(x_{\perp})e^{i({\bf k}- %%@
{\bf k}^{\prime})\cdot {\bf x}}\right]^2
\end{eqnarray}
Here $L_z, L_y$ and $L_x$ are the crystal dimensions over which the plane wave %%@
electron states are normalized and the length of the `infinite' dislocation has %%@
been limited to the size of the crystal along the z direction. $U(x_{\perp})$ %%@
is already defined in Eq. (1) and it does not depend on the z coordinate. So going to cylindrical coordinates, the z %%@
integral is just a delta function. To take the normalization, assume a finite box %%@
of size $L_z$  along the z axis. Therefore $\int_0^\infty d{z} %%@
e^{i({k_z}-{k_z}^{\prime}){z}}\approx L_z \delta_{k_{z}, \, k_{z}^{\prime}}$. With $\delta^2_{k_{z}, \, k_{z}^{\prime}}%%@
=\delta_{k_{z}, \, k_{z}^{\prime}}$, we have
\begin{eqnarray} \label{w- k kprime1}
W_{{\bf k},{\bf k}^{\prime}}={2\pi\over \hbar}
\delta(E_k-E_{k^\prime})\delta_{k_{z}, \, k_{z}^{\prime}}\hspace{2cm}\nonumber\\\times
\left[{L_z\over L_xL_yL_z}{Qe\over2\pi\epsilon}\int d{{\bf x}_{\perp}}K_0({x_{\perp}/\lambda})e^{i({\bf k}_{\perp}- %%@
{\bf k}_{\perp}^{\prime})\cdot {\bf x}_{\perp}}\right]^2.
\end{eqnarray}
The $\theta$ integral in Eq. (\ref{w- k kprime1}) is just the integral representation of the zero-order modified Bessel function of first kind, $J_0$;  $\int_0^{2\pi} d{\bf\theta}\,\exp(i|{\bf k}_{\perp}^{\prime}-{\bf k}_{\perp}|x_{\perp}\cos\theta)=2\pi\,J_0(|{\bf k}_{\perp}- %%@
{\bf k}_{\perp}^{\prime}|\,x_{\perp})$. Further using the identity\cite{tableIntegrals} $\int_0^\infty y\, dy\,K_\nu(ay)J_\nu(by)=%%@
 {b^\nu \over a^\nu(a^2+b^2)}$, here $\nu=0$, the Fourier transform in Eq. (\ref{w- k kprime1}) becomes\cite{look}
\begin{equation}
U(|{\bf k}_{\perp}^{\prime}-{\bf k}_{\perp}|)={Qe \lambda^2 \over  \epsilon %%@
(1+ |{\bf k}_{\perp}^{\prime}-{\bf k}_{\perp}|^2 \lambda^2)}
\end{equation}
 The energy conserving delta function, $\delta(E_k-E_{k^{\prime}})=({\partial E / \partial k} %%@
)^{-1}\delta(k-k^\prime)=({\hbar}^2%%@
{k/m^{\ast}})^{-1}(k/k_\perp)\delta(k_\perp-k_\perp^{\prime})$ due to %%@
$\delta_{k_{z}, \, k_{z}^{\prime}}$ in the summation. Thus, as previously claimed, %%@
both the perpendicular and the parallel components of the electron momenta are %%@
separately conserved. Since, $\Sigma_{\bf k_{\perp}^\prime}\rightarrow %%@
L_xL_y/(2\pi)^2 \int d{\bf k_{\perp}^\prime}$, an overall factor of area remains %%@
in the denominator after the primed momenta have been integrated over. This simply %%@
means that the scattering due to a single charged dislocation is ineffective in a %%@
large sample.\cite{footnote-err} When there are many charged dislocations within %%@
this area which are all parallel, one can simply replace $(L_xL_y)^{-1}$ by $N_d$ %%@
the dislocation density per unit area if the interference terms can be neglected. %%@

\section{Results}
\subsection{Transport Lifetime}
From Eq. (\ref{relaxdef}), the relaxation time in the direction perpendicular %%@
to the dislocation axis is
\begin{equation}\label{relaxation time}
\tau_\perp(k_\perp)={\hbar^3\epsilon^2\over Q^2e^2\lambda^4m^\ast %%@
N_d}[1+(2k_\perp\lambda)^2]^{3/2}
\end{equation}
This is exactly what P\"od\"or had derived $[$Eq. (\ref{podor's expression})$]$ %%@
and $k_\perp=0$ implies a finite $\tau_\perp$ even after our rederivation.
While in three dimensions an electron with $k=0$ is unphysical (there is no %%@
associated phase space), an electron with $k_\perp=0$ and $k_z\neq 0 $ corresponds %%@
to a physical situation. The inconsistency in the final formula results from the %%@
breakdown of the validity of the assumed solution, $\phi_{\bf k}=0$ for %%@
$k_\perp=0$ in Eq. (\ref{ansatz}). This condition is outside the scope of the %%@
present scheme of the solution, which is otherwise consistent.

The anisotropy in $\tau$ necessitates a further angular averaging for a comparison %%@
with any physical quantity associated with a measurement which involves a %%@
thermodynamic distribution of electrons. This transport scattering time is %%@
directly connected to mobility, $\mu=(e/m)\langle\langle \tau \rangle\rangle$, %%@
where $\langle\langle \rangle\rangle$ denote an {\em energy} average, (see below) %%@
over a distribution function of appropriate degeneracy. In a fully degenerate %%@
system, using Eq. (15), this simplifies to %%@
$\langle\langle\tau_{tr}\rangle\rangle=(3/4)\int_0^{\pi}\sin^3\theta \tau_\perp %%@
\,d\theta$.
\subsection{Quantum Scattering Time}
A quantum scattering time, $\tau^{q}_\perp(k_\perp)$ is, by definition, Eq.
(\ref{relaxdef}), but without the $(1-\cos\,\theta)$ factor and may be calculated %%@
similarly. This was done by Jena and Mishra.\cite{jena}
\begin{equation}\label{quantum scattering time}
\tau^{q}_\perp(k_\perp)={\hbar^3\epsilon^2\over Q^2e^2\lambda^4m^\ast %%@
N_d}{[1+(2k_\perp\lambda)^2]^{3/2}\over 1+2(k_\perp\lambda)^2}
\end{equation}
However, the angular dependence of $\tau^q_\perp$ must also to be averaged out.
The meaningful quantity is $\langle %%@
1/\tau^q\rangle=(2/\pi)\int_0^{\pi/2}[\tau^q(\theta)]^{-1}d \theta$ and is often %%@
connected to the finite amplitude and width of the Shubnikov-de Haas or de %%@
Haas-van Alphen oscillations. The quantum scattering time may be looked upon as an %%@
effective `Dingle' temperature, $T_D\sim  (\hbar/ 2\pi k_B) \langle %%@
1/\tau^q\rangle$.

Nevertheless, while comparing Shubnikov amplitudes, the scattering rates are %%@
better calculated between Landau wave functions and with a density of states at %%@
the Fermi level modified by the magnetic field, as was done long back by %%@
Vinokur for the essentially the same problem.\cite{vinokur}
\begin{figure}[h]
\begin{center}
\resizebox{!}{8cm} {\includegraphics{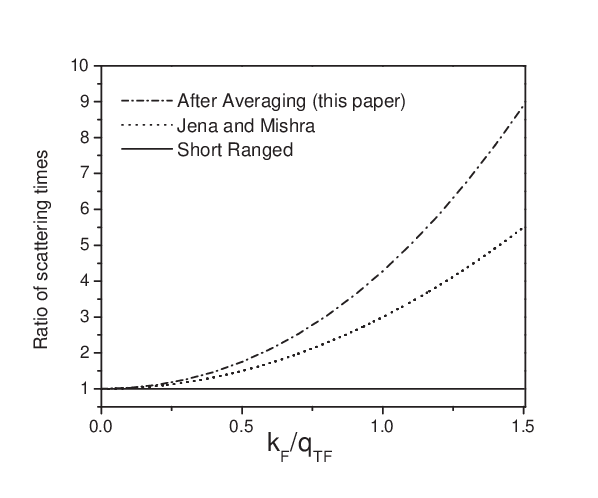}}
\caption{\label{scattering time figure}{ The ratio of dislocation scattering %%@
limited transport and quantum scattering times for a degenerate electron gas. }}
\end{center}
\end{figure}
Furthermore, literature on the connection between scattering times for %%@
dislocations' strain field and de Haas-van Alphen oscillation amplitudes in metals %%@
was a subject of lively debate sometime back. Many parallel interpretations for %%@
level broadening have been suggested.\cite{van alphen controversy}  Some %%@
semiclassical arguments even favour a small angle cutoff. This fact may be %%@
particularly important in two dimensions where it could rescue the quantum %%@
scattering time from a divergence \cite{jena} in a simple and physically %%@
meaningful way, the small angle cutoff $\theta_c$ (in radians) being inversely %%@
proportional to the Landau level index $n$, $\theta_c\simeq \pi/2n$.\cite{cut off %%@
de haas, stormer}

Despite the preceding remarks, the concept of a quantum scattering time finds a %%@
widespread use in literature (for example, references \onlinecite{stormer, jena doped, das sharma stern}). Therefore we have plotted the suitably defined ratio %%@
$\langle 1/\tau^q\rangle \langle\langle\tau_{tr}\rangle\rangle$ of the transport %%@
and quantum scattering times for a three dimensional degenerate carrier gas in  %%@
Fig. \ref{scattering time figure}.
The graph is plotted as a function of the dimensionless parameter, $k_F/q_{TF}$. %%@
$q_{TF}$ is the simple wave vector independent Thomas-Fermi screening function. %%@
The largeness of this ratio is often regarded as a measure of `anisotropy' of %%@
scattering.\cite{hsu walukiewicz} The real space anisotropy of the dislocation %%@
potential is  different from the anisotropy in its Fourier transform, which is %%@
more a measure of the effective range of the potential. An additional averaging %%@
causes the transport to quantum scattering times ratio to be larger than what was %%@
calculated by in reference.\cite{jena}
\subsection{Mobility}
\begin{figure}[h]
\begin{center}
\resizebox{!}{8cm} {\includegraphics{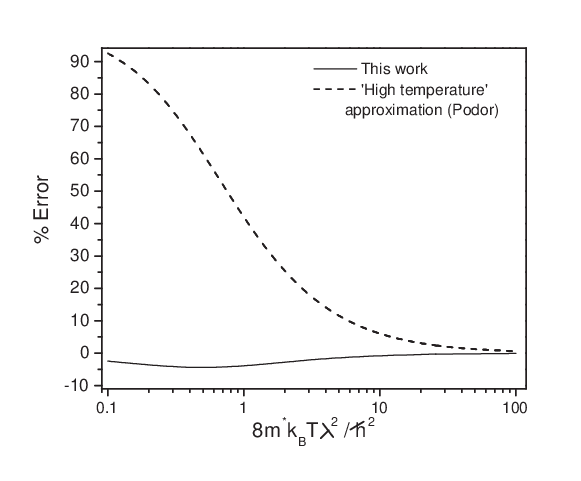}}
\caption{\label{fig:approxs}{ The relative percentage errors %%@
(${\mu_{exact}-\mu_{approx}\over \mu_{exact}}\times 100$) in our formula and %%@
P\"od\"or's approximation with respect to the exact expression evaluated %%@
numerically.
The graphs are plotted as a function of dimensionless parameter ${8m^\ast %%@
k_BT\lambda^2\over \hbar^2}$.}}
\end{center}
\end{figure}

In calculating mobility, the averaging procedure employed by P\"od\"or has been %%@
called `unspecified' and hence it is worked out below.\cite{look} For dislocations %%@
along the z-axis, the current and electric field directions coincide as long as %%@
the measurement is done in the xy-plane. Then, $j_x=ne\langle\langle %%@
v_x\rangle\rangle$ and $\langle\langle v_x\rangle\rangle=\mu_\perp F_x$ where
\begin{equation}
\langle\langle v_x\rangle\rangle={\sum_{\bf k}(f_{0k}+ \phi_{\bf k}) v_x \over %%@
\sum_{\bf k} (f_0+ \phi_{\bf k}) }={\sum_{\bf k}\phi_{\bf k} v_x \over \sum_{\bf %%@
k} f_{0k} }
\end{equation}
Or
\begin{eqnarray}\label{avging definition}
\sigma_{xx}={e^2\hbar^2\over m^{\ast 2}} {2\over (2\pi)^3}\int k_x^2  %%@
\left(-{\partial f_0 \over \partial E}\right) \tau_\perp(k_\perp)d^3k
\label{formula mobility}
\end{eqnarray}
Or
\begin{eqnarray}
\sigma_{xx}={\hbar^5 \epsilon^2\over 2\pi^2m^{\ast 3}Q^2 N_d \lambda^4} \int_0^\pi %%@
d\theta \sin^3\theta  \hspace{2cm}\nonumber\\  \times \int_0^\infty dk k^4 %%@
\left(-{\partial f_0 \over \partial E}\right) %%@
\left[1+(2k\lambda\sin\theta)^2\right]^{3/2}
\end{eqnarray}
The integrals must now be evaluated numerically. Eq. (15) %%@
has the unpleasant feature of depending very strongly on screening length and thus %%@
at low temperatures turns out to be dependent on the model used for the %%@
temperature dependent of carrier concentration and screening. A simple analytic %%@
expression {\it guessed} by interpolating the two integrals ($\int d\theta$ and %%@
$\int dk$) between the two extremes cases, when the first term is much smaller and %%@
when it is much larger than the second term in square brackets in Eq. (15). This is significantly better than P\"od\"or's high %%@
temperature approximation ($k\lambda\sin\theta\gg1$).\cite{seeger}  The relative %%@
percentage errors are plotted in Fig. \ref{fig:approxs}
as a function of the dimensionless parameter ${8m^\ast k_BT\lambda^2\over %%@
\hbar^2}$. It can be seen that this approximation of the integral never deviates %%@
from the numerically calculated exact answer by more than 5\%.

Assuming that the electrons are distributed according to Maxwell-Boltzmann %%@
distribution,
\begin{equation}
\mu_\perp \simeq {2\hbar^3 \epsilon^2\over e\pi^{1/2} m^{\ast ^2} Q^2 N_d %%@
\lambda^4}\left[\pi^{1/3}+
\left( {15\pi\over 8}\right)^{2/ 3}{{8\lambda^2m^{\ast} k_BT}\over\hbar^2} %%@
\right]^{3/2}
\end{equation}
and when the carrier gas is fully degenerate
\begin{equation}
\mu^{\textrm{deg}}_\perp\simeq {3\hbar^5 \epsilon^2\over 4 m^{\ast 3}Q^2 N_d %%@
\lambda_{TF}^4}\left[\left(4/3\right)^{2/3}+\left({5\pi/ %%@
16}\right)^{2/3}4k_F^2\lambda_{TF}^2 \right]^{3/2}
\end{equation}

\subsection{Hall Factor}
In most experiments it is not the drift but the Hall mobility which is measured. %%@
Under the assumption that the scattering rate does not alter in presence of a %%@
magnetic field, B and  when the magnetic field is aligned with the dislocations' %%@
axis, only the in-plane relaxation time comes into the picture. Using the same %%@
line of arguments, it is easy to again establish its existence for arbitrarily %%@
strong non-quantizing magnetic fields. Then, if %%@
$j_x=\sigma_{xx}E_x+\sigma_{xy}E_y$, the Hall scattering factor $r_H$ is defined %%@
as
\begin{equation}\label{hall factor definition}
\\r_H=n\,e{\sigma_{xy}\over B \sigma^2_{xx}}
\end{equation}
where the off-diagonal conductivity, $\sigma_{xy}$, for carriers with parabolic %%@
energy dispersion which are distributed along isotropic constant energy surfaces %%@
is
\begin{eqnarray}\label{off diagonal sigma}
\sigma_{xy}={e^3B\over \hbar^2m^\ast}\int {d^3k\over 4\pi^3}\tau_\perp^2 {\partial %%@
f_0\over \partial E}\left({\partial E\over \partial %%@
k_x}\right)^2\left[1+\left({e\tau_\perp B\over m^\ast}\right)^2\right]^{-1}
\end{eqnarray}
From Eq. (\ref{avging definition}), (\ref{off diagonal sigma}) and (\ref{hall factor definition}) the Hall scattering factor for nondegenerate
carriers at high temperatures (i. e. ${8m^\ast k_BT\lambda^2\over \hbar^2}\gg 1$) approaches a value of $2.07$, obtained by dropping the second term in square brackets in Eq. (\ref{formula mobility}). At lower temperatures, its value is dependent on the model of carrier density and screening but always smaller. The anisotropy in scattering makes the value higher than the Hall factor for ionized impurity scattering which is $1.93$. We see that there can even be a factor of two error in the mobility estimate if the Hall mobility is equated to the drift mobility.
\subsection{Effect of Dislocation Tilt}
Assume that dislocations are all parallel, but now at a longitude $\phi$ and latitude $\theta$ with respect to the z-axis while the measurement is being done in the xy-plane. A unit vector along this dislocation axis is  $\hat{{\bf d}}=
\hat{\textbf{x}} \sin\theta\, \sin\phi + \hat{\textbf{y}} \sin\theta\, \cos\phi + \hat{\textbf{z}}\cos\theta$. Because the electric field is developed only along the direction perpendicular to the dislocations' axis, ${\bf F}_\perp=\rho{\bf j}_\perp=\rho[{\bf j}-({\bf j}\cdot\hat{{\bf d}})\hat{{\bf d}}]$  which yields (with $\cos$ and $\sin$ abbreviated to c and s)
\begin{equation}\label{tensor rotated}
\rho^\prime=\rho\left[\begin{array}{ccc}
 1-\texttt{s}^2\theta \texttt{s}^2\phi&-\texttt{s}^2\theta \texttt{s}\phi
\texttt{c}\phi&-\texttt{c}\theta \texttt{s}\theta \texttt{s}\phi \\
-\texttt{s}^2\theta \texttt{s}\phi \texttt{c}\phi&1-\texttt{s}^2\theta
\texttt{c}^2\phi&-\texttt{c}\theta \texttt{s}\theta \texttt{s}\phi \\
-\texttt{c}\theta \texttt{s}\theta \texttt{s}\phi&-\texttt{c}\theta
\texttt{s}\theta \texttt{s}\phi&1-\texttt{c}^2\theta
\end{array}\right]
\end{equation}
Negative sign in the off diagonals indicates the direction of the electric field developed. Note that tensor $\rho^\prime$ is symmetric, as it should be, to be consistent with Onsager relations. 
\subsection{Angular Distribution of Dislocations}
The extreme anisotropy of the resistivity tensor is usually not seen experimentally. An obvious reason for this that all the dislocations are not parallel to each other. Let us consider the simplest case where the dislocation lines are distributed with a uniform distribution over angles. One can, of course, average over the angles appearing in Eq. (\ref{tensor rotated}).\cite{sondhiemer} This averaging over the angles in the rotated resistivity tensor amounts the use of Matthiessen's rule and will not change the temperature dependence of mobility.

For a better approximation, we again start from the linearized Boltzmann equation, Eq. (\ref{linear-boltzmann}). In the present case, the relaxation time must be isotropic and therefore let the ansatz for the distribution function be
\begin{equation}
\phi(k)=-{\hbar e\over m}{\partial f_{0k}\over\partial E_k}\tau_{iso}(k) {\bf %%@
k\cdot F}
\end{equation}
We shall further assume incoherent scattering such that the scattering rates due to %%@
different dislocation lines add. If the scattering rate due to an $i^{th}$ %%@
dislocation is $W^i_{{\bf k, k^\prime}}$, then the total rate is $\sum_i W^i_{{\bf %%@
k, k^\prime}}$.
\begin{figure}[h]
\begin{center}
\resizebox{!}{7cm} {\includegraphics{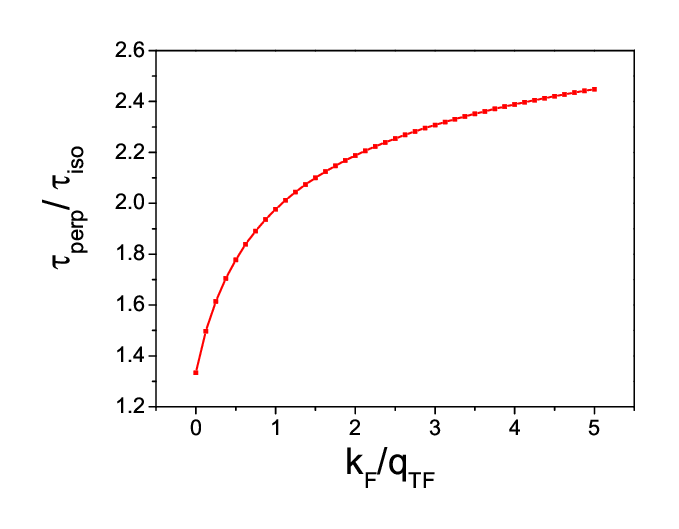}}
\caption{\label{Fig:Ratio_isotropic}{ The ratio of dislocation scattering %%@
limited transport and isotropic scattering times for a degenerate electron gas. }}
\end{center}
\end{figure}
Without loss of generality, one can choose the electron wave vector {\bf k} to be %%@
along the z-axis, ${\bf k}=k\hat{z}$. If the axis of the $i^{th}$ dislocation, %%@
${\bf d}^i$ is at an angle ($\theta, \phi$) with respect to the z-axis, then the %%@
unit vector along the dislocation axis is given by $\hat{d}^i=\sin\theta\sin\phi %%@
\hat{x}+\sin\theta\cos\phi\hat{y}+\cos\theta\hat{z}$.  The component of the wave %%@
vector perpendicular to the dislocation axis is given by
\begin{eqnarray}
{\bf k}^i_\perp={\bf k}-({\bf k\cdot d^i})\hat{d}\,^i \nonumber
\end{eqnarray}
Or
\begin{eqnarray}
{\bf %%@
k}^i_\perp=k\,[(1-\cos^2\theta)\,\hat{z}-\cos\theta\sin\theta\cos\phi\,\hat{y} %%@
\hspace{1cm}\nonumber\\ -\cos\theta\sin\theta\sin\phi\,\hat{x}]
\end{eqnarray}
Substituting back in the Boltzmann equation, we get
\begin{equation}
F_z k_z=-\tau_{iso}(k) {\bf F}\cdot\sum_i \left[{-\bf k}_\perp^i{1\over
\tau(k_\perp^i)}\right]
\end{equation}
Converting the sum into an integral,
\begin{equation}\label{angular 3}
F_zk_z=\tau_{iso}(k)\;{1\over 4\pi}{\bf F}\cdot  \int d\Omega \;{\bf %%@
k}_\perp^i{1\over \tau(k_\perp^i)}
\end{equation}

Since the averaging over the dislocation orientations is equivalent to an %%@
averaging over the electron wave vectors, the expression for the relaxation time %%@
becomes
\begin{equation}\label{iso tau}
{1\over \tau_{iso}(k)}={Q^2e^2\lambda^4 m^\ast N_d \over %%@
2\hbar^3\epsilon^2}\int_0^\pi d\theta{\sin^3\theta \over %%@
[1+4k ^2\lambda^2\sin^2\theta]^{3/2}}
\end{equation}
where the $\phi$ integral has been performed and we have noted that $\int_0^{2\pi} %%@
d\phi\sin\phi =\int_0^{2\pi} d\phi\cos\phi =0$. From here on, it
is straightforward to calculate the isotropic mobility, although
it is best done numerically.\cite{fn} Fig. \ref{Fig:Ratio_isotropic} shows ratio of the perpendicular to the isotropic scattering times for a degenerate electron gas as a function of the dimensionless ratio $k_f/q_{TF}$ where $k_F$ is the Fermi wave vector and $q_{TF}$ is the Fourier transform of the (for example Thomas-Fermi) screening length.
\section{Summary} Within the framework of the conventional transport theory, we %%@
have shown that a relaxation time can be defined for scattering of carriers by %%@
charged dislocations.  Difference between quantum and classical scattering times %%@
was discussed and it was pointed out that the anisotropy necessitates an %%@
appropriate angular averaging. A new approximate formula for mobility was derived %%@
and it was shown be within 5\% of the exact result at all temperatures. The value %%@
of the Hall scattering factor and the effect of dislocation tilt on resistivity %%@
was determined. Finally we derived a new expression for the relaxation time when %%@
the angular orientation of dislocations is isotropic.
\section{Appendix}
\subsection{Variational calculation of mobility}
As a consistency check, let us also consider another method of calculating mobility that avoids the notion of a relaxation time altogether. Following Ziman one can attempt a direct computation of resistivity using thermodynamic arguments and the variational principle.\cite{ziman} In presence of an external electric field, we can write an electron distribution function $f_{\bf k}$ that is shifted from its $k=0$ mean value at equilibrium as
$f_{\bf k}=f_{0{k}}-\Phi_{\bf k}{\partial f_{0k}\over\partial \xi_k}$.  ${\xi_k}$ is the energy gained by the electron from the applied electric field.
$\Phi_{\bf k}$ and $\phi_{\bf k}$ of Section III are obviously related,  $\phi_{\bf k}=-\Phi_{\bf k}{\partial f_{0k}\over\partial \xi_k}$.
The entropy generated per unit time due to current $j$ caused by the applied electric field through the Joule heat dissipated in the material on account of its finite resistivity is $ \dot{S}={\rho j^2/ T} $. Using this thermodynamic argument and the (approximate) expression for entropy in terms of the (perturbed) distribution function, one can write down an expression for resistivity in terms of $\Phi_{\bf k}$ and the scattering rates $W_{\bf k,k^\prime}$.
$$
{\rho}={2\pi^3\over\ k_BT}{{{{\sum_{{\bf k}^\prime}\int d{\bf k}{W_{\bf k,k^\prime}}}
[{\Phi_{\bf k}-\Phi_{\bf k^\prime}}]^2{ f_{0k}}\left[1-f_{0{k}^\prime}\right]}}\over{\left[ {\int d{\bf k}}ev_{k}{\Phi_{k}}{\partial f_{0k}\over\partial \xi_k}\right]}^2}\;\;\;\;\;\;\;(\textrm{A1})
$$
$W_{\bf k,k^\prime}$ are the same as those computed in Section III. According to the variational principle, for any trial function $\Phi_{\bf k}$ the value of the ratio in Eq. (A1) will be greater than or equal to the value of true resistivity, i.e., Eq. (A1) will yield an upper bound of the true resistivity. Thus the computation of resistivity within this framework involves guessing a form for the $\Phi_{\bf k}$ in terms of a variational parameter $s$ and then determining the value of $s$ that minimizes the resistivity computed via Eq. (A1). Writing our trial function\cite{bss}
$$\Phi_{\bf k}=-\tau_{0} {\bf k\cdot \hat{u}}{|k|^s}\;\;\;\;\;\;\;\;\;(\textrm{A2})
$$
where $\hat{u}$ is a unit vector parallel to the applied electric field, we find that Eq. (25) in the high temperature limit yields
$${\rho}={\pi^3 N_{d}Q^{2}\lambda\hbar^{3}\over 128m^*(K_{B}T)^{3}\epsilon^{2}}{\Gamma(s+1)\over\Gamma((s+5)/2)}.\;\;\;\;\;\;\;\;\;(\textrm{A3})$$
In the above equation,  ${\rho}$ is minimum for $s=3$. Hence the calculated mobility using variational principle in high temperature limit is
$${\mu_{var}}={768\sqrt{2}\over \pi^{3/2}}{(K_{B}T)^{3/2}\epsilon^{2}\over m^{*1/2}eN_{d}Q^{2}\lambda}.\;\;\;\;\;\;\;\;\;\;\;\;(\textrm{A4})$$
Comparing this with the high temperature limit of expression for mobility computed in Eq. (15) we find that the two expressions only differ by a  numerical constant with ${\mu_{var}}=1.296\,{\mu_\perp}$. Since the variational principle yields an upper bound on the electrical resistivity, a lower resistivity (high mobility) computed here is probably a better estimate though the small difference in the multiplicative constants appearing in the two expression is experimentally insignificant, especially because $N_d$ is a never known that precisely.


\begin{references}
\bibitem{fn}An explicit plot of the temperature dependence of the
resulting mobility for an isotropic distribution of dislocations is not given here. This is because the screening length depends on the carrier density that has a strong non-universal temperature dependence, especially since the dislocations themselves act as trapping centres for charge carriers.
\end{references}
\end{document}